\documentclass[a4paper,10pt]{article}
\usepackage{graphicx} 
\usepackage{graphics}

\usepackage{amsmath}   
\usepackage{amsfonts}
\usepackage{amssymb}

\newcommand{\dfr}{\widehat{d}} 

\begin{document}
%
%
\def\ov{\over}
\def\l{\left}
\def\r{\right}
\def\beq{\begin{equation}}
\def\eeq{\end{equation}}
\def\d{\partial}
%

\setlength{\oddsidemargin} {1cm}
\setlength{\textwidth}{18cm}
\setlength{\textheight}{23cm}
\title{Derivation of the postulates of quantum mechanics from the first principles of scale relativity}
\author{{\bf Laurent Nottale$^{a,1}$ and Marie-No\"elle C\'el\'erier$^{b,2}$ } \\
{\small LUTH, CNRS, Observatoire de Paris and Paris Diderot University, }\\
{\small 5 place Jules Janssen, 92190 Meudon, France}\\
{\small e-mail: $^a$ laurent.nottale@obspm.fr} \\
{\small $^b$ marie-noelle.celerier@obspm.fr} \\
{\small phone: $^1$ (0)1 45 07 74 03} \\
{\small $^2$ (0)1 45 07 75 38}}
\maketitle

\begin{abstract}

Quantum mechanics is based on a series of postulates which lead to a very good description of the microphysical realm but which have, up to now, not been derived from first principles. In the present work, we suggest such a derivation in the framework of the theory of scale relativity. After having analyzed the actual status of the various postulates, rules and principles that underlie the present axiomatic foundation of quantum mechanics  (in terms of main postulates, secondary rules and derived `principles'), we attempt to provide the reader with an exhaustive view of the matter, by both gathering here results which are already available in the literature, and deriving new ones which complete the postulate list.

\end{abstract}




\section{Introduction}\label{s.i}


Quantum mechanics is a very powerful theory which has led to an accurate description of the micro-physical mechanisms. It is founded on a set of postulates from which the main processes pertaining to its application domain are derived. A challenging issue in physics is therefore to exhibit the underlying principles from which these postulates might emerge.

The theory of scale relativity consists of generalizing to scale transformations the principle of relativity, which has been applied by Einstein to motion laws. It is based on the giving up of the assumption of spacetime coordinate differentiability, which is usually retained as an implicit hypothesis in current physics. Even though this hypothesis can be considered as mostly valid in the classical domain (except possibly at some singularities), it is clearly broken by the quantum-mechanical behavior. It has indeed been pointed out by Feynman (see, e.g., \cite{FH65}) that the typical paths of quantum mechanics are continuous but nondifferentiable. Even more, Abott and Wise \cite{AW81} have observed that these typical paths are of fractal dimension $D_F=2$. This is the reason why we propose that the scale relativity first principles, based on continuity and giving up of the differentiability hypothesis of the coordinate map, be retained as good candidates for the founding of the quantum-mechanical postulates. We want to stress here that, even if coordinate differentiabilty is recovered in the classical domain, nondifferentiability is a fundamental property of the geometry that underlies the quantum realm.

To deal with the scale relativistic construction, one generally begins with a study of pure scale laws, i. e., with the description of the scale dependence of fractal paths at a given point of space (spacetime). Structures are therefore identified, which evolve in a so-called `scale space' that can be described at the different levels of relativistic theories (Galilean, special relativistic, general relativistic) \cite{LN93}. The next step, which we consider here, consists of studying the effects on motion in standard space that are induced by these internal fractal structures.

Scale relativity, when it is applied to microphysics, allows to recover quantum mechanics as a non-classical mechanics on a nondifferentiable, therefore fractal, spacetime \cite{LN93,LN96}. Since we want to limit our study to the basic postulates of nonrelativistic quantum mechanics (first quantization), we focus our attention on fractal power law dilations with a constant fractal dimension $D_F=2$, which means to work in the framework of `Galilean' scale relativity \cite{LN93}.

Now, we come to a rather subtle issue. What is the set of postulates needed to completely describe the quantum-mechanical theory? It is all the more tricky to answer this question that some of the postulates usually presented as such in the literature can be derived from others. We have therefore been led to analyse their status in more detail, then to split the set of `postulates' and `principles' of standard quantum mechanics into three subsets listed in section~\ref{post}: main postulates, secondary ones and derived principles. 

In sections~\ref{cstatef}--\ref{vnbpost}, only the `main postulates' are derived in the framework of scale relativity since the others are mere consequences of these main ones. We explore some miscellaneous related issues in section~\ref{miscel}. Section~\ref{disc} is devoted to the discussion and section~\ref{concl} to the conclusion.

\section{The postulates of quantum mechanics} \label{post}

In the present paper, we examine the postulates listed below. They are formulated within a coordinate realization of the state function, since it is in this representation that their scale relativistic derivation is the most straightforward. Their momentum realization can be obtained by the same Fourier transforms which are used in standard quantum mechanics, as well as the Dirac representation, which is another mathematical formulation of the same theory, can follow from the definition of the wavefunctions as vectors of an Hilbert space upon which act Hermitian operators representing the observables corresponding to classical dynamical quantities.

The set of statements we find in the literature as `postulates' or `principles' can be split into three subsets: the main postulates which cannot be derived from more fundamental ones, the secondary postulates which are often presented as `postulates' but can actually be derived from the main ones, and then statements often called `principles' which are well known to be as mere consequences of the postulates.

\subsection{Main postulates} \label{mainpost}

\begin{enumerate}

\item {\it Complex state function.} Each physical system is described by a state function which determines all can be known about the system. The coordinate realization of this state function, the wavefunction $\psi(r,s,t)$, is an equivalence class of complex functions of all the classical degrees of freedom generically noted $r$, of the time $t$ and of any additional degrees of freedom such as spin $s$ which are considered to be  intrinsically quantum mechanical. Two wave functions represent the same state if they differ only by a phase factor (this part of the `postulate' can be derived from the Born postulate, since, in this interpretation, probabilities are defined by the squared norm of the complex wavefunction and therefore the two wavefunctions differing only by a phase factor represent the same state). The wavefunction has to be finite and single valued throughout position space, and furthermore, it must also be a continuous and continuously differentiable function (but see \cite{Berry1996,Hall2004} on this last point).

\item {\it Schr\"odinger equation.} The time evolution of the wavefunction of a non-relativistic physical system is given by the time-dependent Schr\"odinger equation
\beq
i\hbar \frac{\partial \psi}{\partial t} = \hat{H} \psi \; ,
\label{sch}
\eeq
where the Hamiltonian $\hat{H}$ is a linear Hermitian operator, whose expression is constructed from the correspondence principle.

\item {\it Correspondence principle.} To every dynamical variable of classical mechanics there corresponds in quantum mechanics a linear, Hermitian operator, which, when operating upon the wavefunction associated with a definite value of that observable (the eigenstate associated to a definite eigenvalue), yields this value times the wavefunction. The more common operators occuring in quantum mechanics for a single particle are listed below and are constructed using the position and momentum operators.

\begin{tabular}{lll}
Position & $r (x,y,z)$ & Multiply by $r (x,y,z)$ \\

Momentum & $p(p_x, p_y, p_z)$ & $-i \hbar \nabla$ \\

Kinetic energy & $T=\frac{p^2}{2m}$ & $-\frac{\hbar^2}{2m} \Delta$ \\

Potential energy & $\Phi(r)$ & Multiply by $\Phi(r)$ \\

Total energy & $E = T + \Phi$ & $i \hbar \frac{\partial}{\partial t} = \Phi(r) - \frac{\hbar^2}{2m} \Delta$ \\

Angular momentum & $(l_x,l_y,l_z)$ & $-i\hbar \; r \times \nabla$ \\
\end{tabular}

More generally, the operator associated with the observable $A$ which describes a classically defined physical variable is obtained by replacing in the `properly symmetrized' expression of this variable the above operators for $r$ and $p$. This symmetrization rule is added to ensure that the operators are Hermitian and therefore that the measurement results are real numbers.

However, the symmetrization (or Hermitization) recipe is not unique. As an example, the quantum-mechanical analogue of the classical product $(px)^2$ can be either $(p^2x^2 + x^2p^2)/2$ or $[(xp + px)/2]^2$ \cite{DB54}. The different choices yield corrections of the order of some $\hbar$ power and,  in the end, it is the experiments that decide which is the correct operator. This is clearly one of the main weaknesses of the axiomatic foundation of quantum mechanics \cite{Messiah1959}, since the ambiguity begins with second orders, and therefore concerns the construction of the Hamiltonian itself. 

\item {\it Von Neumann's postulate.} If a measurement of the observable $A$ yields some value $a_i$, the wavefunction of the system just after the measurement is the corresponding eigenstate $\psi_i$ (in the case that $a_i$ is degenerate, the wavefunction is the projection of $\psi$ onto the degenerate subspace).

\item {\it Born's postulate: probabilistic interpretation of the wavefunction.} The squared norm of the wavefunction $|\psi|^2$ is interpreted as the probability of the system of having values $(r,s)$ at time $t$. This interpretation requires that the sum of the contributions $|\psi|^2$ for all values of $(r,s)$ at time $t$ be finite, i. e., the physically acceptable wavefunctions are square integrable. More specifically, if $\psi(r,s,t)$ is the wavefunction of a single particle, $\psi^*(r,s,t)\psi(r,s,t)dr$ is the probability that the particle lies in the volume element $dr$ located at $r$ at time $t$. Because of this interpretation and since the probability of finding a single particle somewhere is $1$, the wavefunction of this particle must fulfill the normalization condition
\beq
\int_{-\infty}^{\infty}\psi^*(r,s,t)\psi(r,s,t)dr = 1 \; .
\label{norm}
\eeq

\end{enumerate}

\subsection{Secondary postulates} \label{secpost}

One can find in the literature other statements which are often presented as `postulates' but which are mere consequences of the above five `main' postulates. We examine below some of them and show how we can derive them from these `main' postulates.

\begin{enumerate}

\item {\it Superposition principle.} Quantum superposition is the application of the superposition principle to quantum mechanics. It states that a linear combination of state functions of a given physical system is a state function of this system. This principle follows from the linearity of the $\hat{H}$ operator in the Schr\"odinger equation, which is therefore a linear second order differential equation to which this principle applies.

\item {\it Eigenvalues and eigenfunctions.} Any measurement of an observable $A$ will give as a result one of the eigenvalues $a$ of the associated operator $\hat{A}$, which satisfy the equation
\beq
\hat{A}\psi = a\psi \; .
\label{eigen}
\eeq

{\bf Proof:} The correspondence principle allows us to associate to every observable a Hermitian operator acting on the wavefunction. Since these operators are Hermitian, their eigenvalues are real numbers, and such is the result of any measurement. This is a sufficient condition to state that any measurement of an observable $A$ will give as a result one of the eigenvalues $a$ of the associated operator $\hat{A}$. Now, we need to prove it is also a necessary condition.

We consider first the Hamiltonian operator, assuming that the classical definition of its kinetic energy part involves only terms which are quadratic in the velocity. The Schr\"odinger equation reads
\beq
i\hbar \frac{\partial \psi}{\partial t} = \hat{H} \psi \; .
\label{schroham}
\eeq
Let us limit ourselves to the case when the potential $\Phi$ is everywhere zero (free particle). For simplification purpose, we ignore also here the dependence on $s$ of the state function. Equation~(\ref{schroham}) becomes
\beq
i\hbar \frac{\partial}{\partial t}\psi(r,t) = - \frac{\hbar ^2}{2m} \Delta \psi(r,t) \; .
\label{schrozeropot}
\eeq
This differential equation has solutions of the kind
\beq
\psi(r,t) = C \rm{e}^{i(k.r - \omega t)} \; ,
\label{planewsol}
\eeq
where $C$ is a constant and $k$ and $\omega$ verify
\beq
\omega = \frac{\hbar k^2}{2m} \; .
\label{debroglie}
\eeq
We apply now to the expression of $\psi$ in equation~(\ref{planewsol}) the operators $\hat{P} = -i \hbar \nabla$ and $\hat{E} = i\hbar \partial / \partial t$ and we obtain
\beq
\hat{P} \psi(r,t)= -i \hbar \nabla \psi(r,t) = \hbar k \psi(r,t) \; ,
\label{hatpplanew}
\eeq
and
\beq
\hat{E} \psi(r,t) = i\hbar \frac{\partial}{\partial t}\psi(r,t) = \hbar \omega \psi(r,t) \; .
\label{hateplanew}
\eeq
An inspection of equations~(\ref{debroglie})--(\ref{hateplanew}) shows that the eigenvalues $\hbar k$ of $\hat{P}$ and $\hbar \omega$ of $\hat{E}$ are related in the same way as the momentum $p$ and the energy $E$ in classical physics, i. e., $E = p^2/2m$. Owing to the correspondence principle, we can therefore assimilate $\hbar k$ to a momentum $p$ and $\hbar \omega$ to an energy $E$, thus recovering the de Broglie relation $p = \hbar k$ and the Einstein relation $E = \hbar \omega$. Moreover, since equations~(\ref{hatpplanew}) and (\ref{hateplanew}) can be rewritten
\beq
\hat{P} \psi(r,t)= p \psi(r,t) \; 
\label{eigenp}
\eeq
and
\beq
\hat{E} \psi(r,t) = E \psi(r,t) \; ,
\label{eigene}
\eeq 
we have shown that, in the case where the classical definition of the (free) Hamiltonian writes $H = P^2/2m$, the measurement results of the $P$ and $E$ observables are eigenvalues of the corresponding Hermitian operators.

As regards the position operator, this property is straightforward since the application of the correspondence principle implies that to $r$ corresponds the operator `multiply by $r$'. We therefore readily obtain
\beq
\hat{R} \psi = r \psi \; ,
\label{eigenr}
\eeq
which implies that $r$ is actually the eigenvalue of $\hat{R}$ obtained when measuring the position.

Implementing the correspondence principle, these results can be easily generalized to all other observables which are functions of $r$, $p$ and $E$.  

\item {\it Expectation value.} For a system described by a normalized wave function $\psi$, the expectation value of an observable $A$ is given by
\beq
\langle A \rangle = \int_{-\infty}^\infty \psi^* \hat{A} \psi \, dr \; .
\label{expect}
\eeq
This statement follows from the probabilistic interpretation attached to $\psi$, i. e., from Born's postulate (for a demonstration see, e. g., \cite{BJ00}).

\item {\it Expansion in eigenfunctions.} The set of eigenfunctions of an operator $\hat{A}$ forms a complete set of linearly independent functions. Therefore, an arbitrary state $\psi$ can be expanded in the complete set of eigenfunctions of $\hat{A}$ $(\hat{A}\psi_n = a_n \psi_n)$, i. e., as
\beq
\psi = {\sum_n} c_n \psi_n \; ,
\label{exp}
\eeq
where the sum may go to infinity. For the case where the eigenvalue spectrum is discrete and non-degenerate and where the system is in the normalized state $\psi$, the probability of obtaining as a result of a measurement of $A$ the eigenvalue $a_n$ is $|c_n|^2$. This statement can be straightforwardly generalized to the degenerate and continuous spectrum cases.

Another more general expression of this postulate is `an arbitrary wave function can be expanded in a complete orthonormal set of eigenfunctions $\psi_n$ of a set of commuting operators $A_n$'. It writes
\beq
\psi = \sum_n c_n \psi _n \; ,
\label{exps}
\eeq
while the statement of orthonormality is
\beq
\sum_s \int \psi^*_n(r,s,t)\psi _m(r,s,t)dr = \delta_{nm} \; ,
\label{orth}
\eeq
where $\delta_{nm}$ is the Kronecker symbol.

{\bf Proof:} Hermitian operators are known to exhibit the two following properties: (i) two eigenvectors of a Hermitian operator corresponding to two different eigenvalues are orthogonal, (ii) in an Hilbert space with finite dimension $N$, a Hermitian operator always possesses $N$ eigenvectors that are linearly independent. This implies that, in such a finite dimensional space, it is always possible to construct a base with the eigenvectors of a Hermitian operator and to expand any wavefunction in this base. However, when the Hilbert space is infinite this is not necessarily the case any more. This is the reason why one introduces the observable tool. An Hermitian operator is defined as an observable if its set of orthonormal eigenvectors is complete, i. e., determines a complete base for the Hilbert space (see, e. g., \cite{CT77}). 

The probabilistic interpretation attached to the wavefunction (Born's postulate) implies that, for a system described by a normalized wavefunction $\psi$, the expectation value of an observable $A$ is given by
\beq
\langle A \rangle = \int_{-\infty}^\infty \psi^* \hat{A} \psi \, dr = \langle \psi|\hat{A}|\psi \rangle \; .
\label{expect2}
\eeq
Expanding $\psi$ in a complete eigenfunction set of $A$ (or in a complete eigenfunction set of commuting operators), $\psi = \sum_n c_n \psi_n$, where the $c_n$'s are complex numbers, gives
\beq
\langle A \rangle = \sum_m \sum_n c_m^* c_n \langle \psi_m|\hat{A}|\psi_n \rangle = \sum_m \sum_n c_m^* c_n a_n \langle \psi_m|\psi_n \rangle = \sum_n |c_n|^2 a_n \; ,
\label{expect3}
\eeq
since, from orthonormality, $\langle \psi_m|\psi_n \rangle = \delta_{mn}$.

Assuming that $\psi$ is normalized, i. e., $\langle \psi|\psi \rangle = 1$, we can write
\beq
\sum_n |c_n|^2 = 1 \; .
\label{normcn}
\eeq
From the eigenvalue secondary postulate, the results of measurements of an observable $A$ are the eigenvalues $a_n$ of $\hat{A}$. Since the average value obtained from series of measurements of a large number of identically prepared systems, i. e., all in the same state $\psi$, is the expectation value $\langle A \rangle$, we are led, following the Born postulate, to identify the quantity
\beq
P_n = |c_n|^2 = |\langle \psi_n|\psi \rangle |^2
\label{proban}
\eeq
with the probability that, in a given measurement of $A$, the value $a_n$ would be obtained \cite{BJ00}.

To derive these results we have implicitly included the degeneracy index in the summations. A generalization to a degenerate set of eigenvalues is straightforward, and such is the generalization to a continuous spectrum \cite{BJ00}.

It is worth stressing here that this secondary postulate is readily derived from Born's postulate and the superposition principle which itself is a mere consequence of the linearity of the $\hat{H}$ operator and of the Schr\"odinger equation. Therefore it is not an actual founding postulate even if it is often presented as such in the literature (see, e. g., \cite{CT77}).

\item {\it Probability conservation.} The probability conservation is a consequence of the Hermitian property of $\hat{H}$ \cite{CT77}. This property first implies that the norm of the state function is time independent and it also implies a local probability conservation which can be written (e. g., for a single particle without spin and with normalized wavefunction $\psi$) as
\beq
\frac{\partial}{\partial t} \rho(r,t) + \rm{div} J(r,t) = 0 \; ,
\label{locpcons}
\eeq
where
\beq
J(r,t) = \frac{1}{m} {\rm Re} \left[\psi^*\left(\frac{\hbar}{i} \nabla \psi\right)\right] \; .
\eeq

\item {\it Reduction of the wave packet or projection hypothesis.} This statement does not need to be postulated since it can be deduced from other postulates (see, e. g., \cite{RB89}). It is actually implicitly contained in von Neumann's postulate.

\subsection{Derived principles} \label{derprinc}

\item {\it Heisenberg's uncertainty principle.} If $P$ and $Q$ are two conjugate observables such that their commutator equals $i\hbar$, it is easy to show that their standard deviations $\Delta P$ and $\Delta Q$ satisfy the relation
\beq
\Delta P  \, \Delta Q \geq \frac{\hbar}{2} \; ,
\label{heis}
\eeq
whatever the state function of the system \cite{CT77,BJ00}. This applies to any couple of linear (but not necessarily Hermitian) operators and, in particular, to the couples of conjugate variables: position and momentum, time and energy. Moreover, generalized Heisenberg relations can be established for any couple of variables \cite{Finkel1987}.

\item {\it The spin-statistic theorem.}  When a system is composed of many identical particles, its physical states can only be described by state functions which are either completely antisymmetric (fermions) or completely symmetric (bosons) with respect to permutations of these particles, or, identically, by wavefunctions that change sign in a spatial reflection (fermions) or that remain unchanged in such a transformation (bosons). All half-spin particles are fermions and all integer-spin particles are bosons.
 
Demonstrations of this theorem have been proposed in the framework of field quantum theory as originating from very general assumptions. The usual proof can be summarized as follows: one first shows that if one quantizes fermionic fields (which are related to half-integer spin particles) with anticommutators one gets a consistent theory, while if one uses commutators, it is not the case; the exact opposite happens with bosonic fields (which correspond to integer spin particles), one has to quantize them with commutators instead of anticommutators, otherwise one gets an inconsistent theory. Then, one shows that the (anti)commutators are related to the (anti)symmetry of the wavefunctions in the exchange of two particles. However, this proof has been claimed to be incomplete \cite{CT77} but more complete ones have been subsequently proposed \cite{RF86,MY00}.

\item {\it The Pauli exclusion principle.} Two identical fermions cannot be in the same quantum state. This is a mere consequence of the spin-statistic theorem.

\end{enumerate}

Now, we come to the derivation from the scale relativity first principles of the five `main' postulates listed in section~\ref{mainpost}. The other rules and principles which have been themselves shown to derive from these main postulates will be then automatically established.

\section{The complex state function} \label{cstatef}

For sake of simplification, we consider states which are only dependent on the position $r$ and the time $t$. It can however be shown that the spin, which is an intrinsic property of quantum mechanical systems, also naturally arises in the scale relativistic framework \cite{CN04,CN06}, so that the various derivations can easily be generalized to more complex states.

Scale relativity extends the founding stones of physics by giving up the hypothesis of spacetime differentiability, while retaining the continuity assumption. The manifolds can be ${\cal C}^0$ instead of at least ${\cal C}^2$ in Einstein's general relativity. However, the coordinate transformations can be or cannot be differentiable and the scale relativity theory includes therefore general relativity and classical mechanics since, as we shall recall, its basic description of infinitesimal displacements in spacetime is made in terms of the sum of a classical (differentiable) part and of a fractal (nondifferentiable) stochastic part. It must also be emphasized that, here, the nondifferentiability means that one loses derivatives in their usual sense, but that one keeps the ability to define differential elements thanks to continuity.

One of the main consequences of the nondifferentiable geometry, in the simplest case corresponding to non-relativistic motion, is that the concept of velocity (which can be redefined in terms of divergent fractal functions) becomes two valued, implying a two valuedness of the Lagrange function and therefore of the action. Finally, the wavefunction is defined as a re-expression of the action, so that it is also two valued and therefore complex  \cite{LN93,LN96,CN04}.

We briefly recall here the main steps of the construction of this wavefunction from an extension to scales of the relativity and geodesic principles, limiting our study to the simplest case of a nondifferentiable and therefore fractal 3-space, where the time $t$ is not fractal but is retained as a curvilinear parameter. The generalization of this method to the four-dimensional spacetime with the proper time $s$ as a curvilinear parameter allows one to recover spinors and the Dirac and Pauli equations \cite{CN04,CN06}, but we shall not detail this case in the present paper.

A first consequence of the giving up of the differentiability assumption can be derived from the following fundamental theorem: {\it A continuous and non-differentiable (or almost nowhere differentiable) curve is explicitly scale-dependent, and its length tends to infinity when the scale interval tends to zero} \cite{LN93,LN96,BC00}. Since the generalization of this theorem to three (four) dimensions is straightforward, we can say that a continuous and nondifferentiable space(time) is fractal, under the general meaning of scale divergence \cite{BM82}.

Another consequence of the nondifferentiable nature of space is the breaking of local differential time reflection invariance. The derivative with respect to the time $t$ of a function $f$ can be written twofold
\beq
\frac{df}{dt} = \lim_{dt \rightarrow 0}\frac{f(t+dt) - f(t)}{dt} = 
\lim_{dt \rightarrow 0}\frac{f(t) - f(t-dt)}{dt} \; .
\label{derbis}
\eeq
Both definitions are equivalent in the differentiable case. In the 
nondifferentiable situation, these definitions fail, since the limits are no longer defined. In the framework of scale relativity, the physics is related to the behavior of the function during the `zoom' operation on the time resolution $\delta t$, here identified with the differential element $dt$, which is considered as an independent variable \cite{CN04}. The standard function $f(t)$ is therefore replaced by a fractal function $f(t, dt)$, explicitly dependent on the time resolution interval, whose derivative is undefined only at the unobservable limit $dt \rightarrow 0$. As a consequence, one is led to define the two derivatives of a fractal function as explicit functions of the two variables $t$ and $dt$,
\beq
f'_+(t,dt) = \l\{ \frac{f(t+dt,dt)-f(t,dt)}{dt} \r\}_{dt \to 0} \; ,
\label{derp}
\eeq
\beq
f'_-(t,dt) = \l\{ \frac{f(t,dt)-f(t-dt,dt)}{dt}\r\}_{dt \to 0} \; ,
\label{derm}
\eeq
where the only difference with the standard definition is that we still consider $dt$ tending toward zero , but without going to the limit which is now undefined. In other words, instead of considering only what happens at the limit $dt=0$ as in the standard calculus, we consider the full and detailed way by which the function changes when $dt$ becomes smaller and smaller. The two derivatives are transformed one into the other by the transformation $dt \leftrightarrow -dt$ (local differential time reflection), which is an implicit discrete symmetry of differentiable physics.

When one applies the above reasoning to the three space coordinates, generically denoted $X$, one sees that the velocity 
\beq
V = {\frac{dX}{dt}}= \lim_{dt \rightarrow 0} \frac{X(t+dt) - X(t)}{dt}
\label{eq.6}
\eeq
is undefined. But it can be redefined in the manner of equations~(\ref{derp}) and (\ref{derm}) as a couple of explicitly scale-dependent fractal functions $V_\pm(t,dt)$, whose existence is now ensured for any non-zero $dt \to 0$, except at the unphysical limit $dt= 0$.

The scale dependence of these functions suggests that the standard equations of physics be completed by new differential equations of scale. With this aim in view, one considers first the velocity at a given time $t$, which amounts to the simplified case of a mere fractal function $V(dt)$. Then one writes the simplest possible equation for the variation of the velocity in terms of the scale variable $dt$, as a first order differential equation $dV/d \ln dt=\beta(V)$ \cite{LN93,LN96}. Then Taylor expanding it, using the fact that $V<1$ (in units $c=1$), one obtains the solution as a sum of two terms, namely, a scale-independent, differentiable, `classical' part and a divergent, explicitly scale-dependent, non-differentiable `fractal' part \cite{LN97A},
\beq
V = v + w = v \; \left[1 + a \left(\frac{\tau}{dt}\right)^{1-1/D_F}\right],
\label{eq.7}
\eeq
where $D_F$ is the fractal dimension of the path. This relation can be readily generalized to coefficients $v$ and $a$ that are functions of other variables, in particular of the space and time coordinates.

The transition scale $\tau$ yields two distinct behaviors of the velocity depending on the resolution at which it is considered, since $V \approx v$ when $dt \gg \tau$ and $V \approx w$ when $dt \ll \tau$. In the case when this description holds for a quantum particle of mass $m$, $\tau$ is related to the de Broglie scale of the system ($\tau=\hbar/mv^2$) and the explicit `fractal' domain with the quantum domain. 

Now one does not deal here with a definite fractal function, for which the function $a$ could be defined, but with fractal paths which are characterized by the only statement that they are the geodesics of a nondifferentiable space (we shall recall in what follows how the equation for these geodesics can be written). Such a space can be described as being everywhere singular \cite{LN93}, so that there is a full loss of information about position, angles and time along the paths, at all scales $dt< \tau$. In other words, this means that the geodesics of a nondifferentiable space are no longer deterministic. Such a loss of determinism, which is one of the main features of the quantum-mechanical realm, is not set here as a founding stone of the theory, but it is obtained as a manifestation on the paths of the nondifferentiable geometry of spacetime.

As a consequence of this loss of determinism, the scaled fluctuation $a$ is no longer defined. Therefore, it is described by a dimensionless stochastic variable which is normalized according to $\langle a \rangle = 0$ and $\langle a^2\rangle = 1$, where the mean is taken on the probability distribution of this variable. As we shall see, the subsequent development of the theory and its results do not depend at all on the choice of this probability distribution. This representation is therefore fully general and amounts simply to defining a proper normalization of the variables.

The total loss of information at each time-step and at each scale of the elementary displacements in the nondifferentiable space has also another fundamental consequence. It leads to a value $D_F = 2$ of the fractal dimension, which is known to be the Markovian value free of correlations or anticorrelations \cite{BM82}. Moreover one can show that $D_F = 2$ plays the role of a critical dimension, since for other values one still obtains a Schr\"odinger form for the equation of dynamics, but which keeps an explicit scale dependence \cite{LN96}. Therefore, low energy quantum physics can be identified with the case where the fractal dimension of space paths is $2$ (and relativistic quantum physics with the same fractal dimension for spacetime paths \cite{LN93,Ord1983}), in agreement with Feynman's characterization of the asymptotic scaling quantum domain as $w \propto (dt/\tau)^{-1/2}$ \cite{FH65}.

However, this is not the last word, since the loss of determinism on a fractal space has still another fundamental consequence. Indeed, one can prove that the fractality and nondifferentiability of space imply that there is an infinity of fractal geodesics relating any couple of its points (or starting from any point) \cite{LN93,Cresson2001}, and this at all scales. Knowing that the infinite resolution $\delta x=0, \delta t=0$ is considered as devoid of physical meaning in the scale relativity approach, it becomes therefore impossible and physically meaningless to define or to select only one geodesical line. Whatever the subsample considered for whatever small subset of space, time, or any other variable, the physical object (which, as we shall see, the wavefunction describes) remains a bundle made of an infinity of fractal geodesics. 

This is in accordance with Feynman's path integral formulation of quantum mechanics and with his characterization of the typical paths of a quantum-mechanical particle as being in infinite number, nondifferentiable and fractal (in modern words).  But this also allows one to go farther, and to identify the `particles' themselves and their `internal' properties with the geometrical properties of the geodesic bundle corresponding to their state, according to the various conservative quantities that define them \cite{LN93,LN96}. In other words, in the scale relativity framework, one no longer needs to consider `particles' that would follow trajectories described as geodesics of a given spacetime geometry (as in general relativity): the so-called `particles' (i.e., a set of wavefunctions characterized by quantized conservative quantities which possibly change their repartition during the time evolution) are identified with the geodesics themselves, and are therefore considered as pure geometric and relative quantities, that do not exist in an absolute way.

In order to account for the infinity of geodesics in the bundle, for their fractality and for the two valuedness of the derivative which all come from the nondifferentiable geometry of the spacetime continuum, one therefore adopts a generalized statistical fluid-like description where, instead of a classical deterministic velocity $V(t)$ or of a classical fluid velocity field $V[x(t),t]$, one uses a doublet of fractal functions of space coordinates and time which are also explicit functions of the resolution interval $dt$, namely, $V_+[x(t,dt),t,dt]$ and $V_-[x(t,dt),t,dt]$. 

According to equation~(\ref{eq.7}), which can be easily generalized to this case, these two velocity fields can be in turn decomposed in terms of a `classical' part, which is differentiable and independent of resolution, and of a `fractal' part, $V_{\pm}[x(t,dt),t,dt]=v_{\pm}[x(t),t]+w_{\pm}[x(t,dt),t,dt]$. There is no reason {\it a priori} for the two classical velocity fields to be equal.

We thus see that, while, in standard mechanics, the concept of velocity was one valued,  two velocity fields must be introduced even when going back to the classical domain. Therefore, reversing the sign of the time differential element, $v _{+}$ becomes $v _{- }$. A natural solution to this problem is to consider both $(dt > 0)$ and  $(dt < 0) $ processes on the same footing and to combine them in a unique twin process in terms of which the invariance by reflection is recovered. The information needed to describe the system is therefore doubled with respect to the standard description. A simple and natural way to account for this doubling is to use complex numbers and the complex product to define a complex derivative operator \cite{LN93}
\beq
\frac{\dfr}{dt} = \frac{1}{2} \left( \frac{d_+}{dt} + \frac{d_-}{dt} \right) 
- \frac{i}{ 2} \left(\frac{d_+}{dt} - \frac{d_-}{dt}\right) \; .
\label{derop}
\eeq
A detailed justification of the choice of complex numbers to account for this two valuedness has been given in \cite{CN04}.

Now, following the expression obtained above for the velocity, one may write the elementary displacements for both processes as 

\beq
dX_\pm=v_{\pm}[x(t),t]\,dt +w_{\pm}[x(t,dt),t,dt]\,dt \; ,
\label{diffel}
\eeq

i. e., as the sum of a `classical' part, $dx_\pm = v_\pm\;dt$, and of a stochastic  `fractal' part, $d\xi_\pm$, fluctuating about this classical part and such that $\langle d\xi_\pm\rangle = 0$ and that
\beq
\langle d\xi_{\pm i}\;d\xi_{\pm j}\rangle = \pm 2 \; {\cal D} \; \delta _{ij} \; dt \; ,
\eeq
with $i,j=x,y,z$, namely,
\beq
dX_+(t) = v_+\;dt + d\xi_+(t) \quad \quad dX_-(t) = v_-\;dt + d\xi_-(t) \; .
\label{elemdispl}
\eeq
Applying the complex derivative operator of equation~(\ref{derop}) to this position vector yields therefore a complex velocity. In the first works on the scale relativity derivation of the Schr\"odinger equation \cite{LN93,LN96}, only the classical part was considered for this definition, owing to the fact that the fractal part is of zero mean. In more recent works \cite{LN99,LN07,LN07A}, a full complex velocity field has been defined, 
\begin{equation}
\widetilde{\cal V}={ \frac{\dfr}{dt} X(t)=\cal V} + {\cal W} = \left(\frac{v_{+}+v_{-}}{2} -i \, \frac{v_{+}-v_{-}}{2} \right) + \left(\frac{w_{+}+w_{-}}{2}-i \, \frac{w_{+}-w_{-}}{2}\right),
\label{covel}
\end{equation} 
which also contains the fractal divergent part. As we shall see, the various elements of the subsequent derivation remain valid in this case (in terms of fractal functions) and one obtains in the end the same Schr\"odinger equation \cite{LN07,LN07A}, but which now allows nondifferentiable and fractal solutions, in agreement with \cite{Berry1996,Hall2004}.

As we shall show in section~\ref{schroeq}, the transition from classical (differentiable) mechanics to the scale relativistic framework is implemented by replacing the standard time derivative $d/dt$ by $\dfr/dt$ \cite{LN93,CN04} (being aware, in this replacement, that it is a linear combination of first and second order derivatives (see equation~(\ref{covderop})), in particular when using the Leibniz rule \cite{JCP99,LN04}). This means that $\dfr/dt$ plays the role of a `covariant derivative operator', i. e., of a tool that preserves the form invariance of the equations.

The next step consists of constructing the wavefunction by generalizing the standard classical mechanics using this covariance (see section~\ref{lagrang}). A complex Lagrange function and the corresponding complex action are obtained from the classical Lagrange function $L (x, v, t)$ and the classical action $S$ by replacing $d/dt$ by $\dfr/dt$. The stationary action principle applied to this complex action yields generalized Euler-Lagrange equations \cite{LN93,CN04}. 

A complex wavefunction $\psi$ is then defined as another expression for the complex action ${\cal S}$ \cite{LN96,CN04}, namely,
\beq
\psi^{-1} \nabla \psi = {\frac{i}{S_0}} \nabla {\cal S} \; ,
\label{cwfunct}
\eeq
where $S_0$ is a constant which is introduced for dimensional reasons, since ${\cal S}$ has the dimension of an angular momentum. The justification of the interpretation of this complex function as a wavefunction will be given in the following sections of this paper, where it will be shown to own all the properties verified by the wavefunction of quantum mechanics.

\section{The Schr\"odinger equation} \label{schroeq}

The next step is now the derivation of the Schr\"odinger equation. While, in the axiomatic foundation of quantum mechanics, it is often set as the last postulate, we shall see that it plays, in the scale relativity approach, a key role for the derivation of the other postulates. In such a geometric approach, it is derived as a geodesic equation. Only the main steps of the reasoning will be given here since the detailed derivations can be found in the literature \cite{LN93,LN96,CN04,LN99,LN07}.

\subsection{Complex time-derivative operator} \label{tderop}

A complex derivative operator, $\dfr/dt$, has been defined by equation~(\ref{derop}) which, applied to the position vector, yields a complex velocity (equation(\ref{covel})).

In the case of a fractal dimension $D_F=2$ considered here, the total derivative of a function $f$ should be written up to the second order partial derivatives,
\beq
\frac{df} {dt} = \frac{\partial f}{\partial t} + \frac{\partial f }{\partial x_i} \frac{dX_i}{dt} + 
\frac{1}{2} \frac{\partial ^2 f}{\partial x_i \partial x_j} \frac{dX_i dX_j} 
{dt} \; .
\label{totder}
\eeq
Actually, if only the `classical part' of this expression is now considered, one can write $\langle dX \rangle=dx$, so that the second term reduces to $v. \nabla f$, and so that $dX_i dX_j /dt$, which is infinitesimal in the standard differentiable case, reduces to $\langle d \xi_i\; d\xi_j \rangle  /dt$. Therefore the last term of the `classical part' of equation~(\ref{totder}) amounts to a Laplacian, and the two-valued derivative reads  
\beq
\frac{d_{\pm}f} {dt} = \left(\frac{\partial}  {\partial t} + v_\pm . \nabla 
\pm {\cal D} \Delta\right) f \; .
\label{cptotder}
\eeq
Substituting equations~(\ref{cptotder}) into equation~(\ref{derop}), one finally 
obtains the expression for the complex time-derivative operator \cite{LN93}
\beq
\frac{\dfr}{dt} = \frac{\partial}{\partial t} + {\cal V}. \nabla - 
i {\cal D} \Delta \; .
\label{covderop}
\eeq
This operator $\dfr/dt$ plays the role of a `covariant derivative operator', namely, it is used to write the fundamental equation of dynamics under the same form it has in the classical and differentiable case. 

Under its above form, the covariant derivative operator is not itself fully covariant since it involves second order derivative terms while it is a first order time derivative.  These second order terms imply that the Leibniz rule for a product is no longer the first order Leibniz rule \cite{JCP99}, but a linear combination of the first and second order rules \cite{LN04}. 

The strong covariance can be fully implemented by introducing new tools allowing us to keep the form of the first order Leibniz rule, despite the presence of the second order derivatives \cite{JCP99,LN04}. To this purpose, one defines a complex velocity operator \cite{LN04}
\beq
\widehat{\cal V}= {\cal V}-i {\cal D} \nabla.
\eeq
When it is written in function of this operator, the covariant derivative recovers the standard first order form of the expression of a total derivative in terms of partial derivatives, namely,
\beq
\frac{\dfr}{dt}= \frac{\d}{\d t} + \widehat{\cal V}. \nabla.
\label{OK}
\eeq
More generally, one may define, for any function $f$, the operator \cite{LN04}
\beq
\widehat{\frac{\dfr f}{dt}}= \frac{\dfr f}{\d t} -i {\cal D}\,  \nabla f. \nabla,
\eeq
in terms of which the covariant derivatives of a product and of composed functions keep their first order forms.

The covariant derivative operator given by equation~(\ref{OK}) is not yet fully general, since it acts on the classical parts of the various physical quantities, in particular of the velocity field. But, by defining a full velocity operator $ \widehat{\widetilde{\cal V}} =\widetilde{\cal V}-i {\cal D} \nabla$, it can be generalized to the full complex velocity field as
\begin{equation}
\frac{\widehat{d}}{dt} =  \frac{\d}{\d t} + \widehat{\widetilde{\cal V}}. \nabla,
\end{equation}
plus infinitesimal terms that vanish when $dt \to 0$ \cite{LN99,LN07}.

\subsection{Lagrangian approach} \label{lagrang}

Standard classical mechanics can now be generalized using this covariance. Considering motion in standard space, let us first retain only the `classical parts' of the variables, which are differentiable and independent of resolutions. The effects of the internal nondifferentiable structures are contained in the covariant derivative. We assume that the `classical part' of the mechanical system under consideration can be characterized by a Lagrange function ${\cal L} (x, {\cal V}, t)$ (where $x$ and $\cal V$ are three dimensional), that keeps the usual form in terms of the complex velocity ${\cal V}$ and from which a complex action ${\cal S}$ is defined as
\beq
{\cal S} = \int_{t_1}^{t_2} {\cal L} (x, {\cal V}, t) dt \; .
\label{complact}
\eeq
Implementing a stationary action principle, generalized Euler-Lagrange equations are obtained, that read \cite{LN93,CN04}
\beq
\frac{\dfr}{dt} \, \frac{\partial L}{\partial {\cal V}} = \frac{\partial L}{\partial x} \; .
\label{geneuleq}
\eeq
Thanks to the transformation $d/dt \rightarrow \dfr/dt$, they exhibit exactly their standard classical form. This reinforces the identification of the scale relativity tool with a `quantum-covariant' representation.

One may now use Noether's theorem and construct the conservative quantities which find their origin in the spacetime symmetries. From the homogeneity of standard space a generalized complex momentum is defined as
\beq
{\cal P} = \frac{\partial {\cal L}}{\partial {\cal V}} \; .
\label{gencomplmom1}
\eeq
Considering the action as a functional of the upper limit of integration in the action integral, the variation of the action from a trajectory to another nearby trajectory yields a generalization of another well-known relation of standard mechanics,
\beq
{\cal P} = \nabla {\cal S} \; .
\label{gencomplmom2}
\eeq
And from the uniformity of time,  the energy (i. e., in terms of the coordinates and momenta, the Hamiltonian of the system) can be obtained,
\beq
{\cal H} = -\frac{\partial {\cal S}}{\partial t} \; .
\label{hamilt}
\eeq
In the general case where the Lagrange function writes (`particle' in a scalar potential)
\beq
{\cal L} = \frac{1}{2}\, m\, {\cal V}^2-\Phi \; ,
\label{lagr}
\eeq
the complex momentum ${\cal P}=\partial  {\cal L}/ \partial {\cal V}$ keeps its standard form, in terms of the complex velocity field,
\beq
{\cal P}= m {\cal V} \; ,
\label{momenmv}
\eeq
and the complex velocity follows as
\beq
{\cal V} = \nabla {\cal S}/ m \; .
\label{vasgrads}
\eeq
Concerning the generalized energy, its expression involves an additional term \cite{LN97A,LN98--LN04}, namely, the Hamilton function writes
\beq
{\cal H}= \frac{1}{2} m ({\cal V}^{2}-2i \, {\cal D} \, {\rm div} {\cal V}) \; .
\eeq
The origin of the additional energy is actually a direct consequence of the second order term in the covariant derivative \cite{LN98,JCP99}. 

For the Lagrangian of equation~(\ref{lagr}), the Euler-Lagrange equations keep the form of Newton's fundamental equations of dynamics, 
\beq
m\, \frac{\dfr}{dt} {\cal V}= - \nabla \Phi \; ,
\label{neweq}
\eeq
which is now written in terms of complex variables and of a complex time-derivative operator.

In the case where there is no external field, the strong covariance is explicit, since equation~(\ref{neweq}) takes the form of the equation of inertial motion, i. e., of a geodesic equation,
\beq
\frac{\dfr \;{\cal V}}{dt} = 0 \; .
\label{geoeqin}
\eeq
This is one of the main results of the scale relativity theory, since it means that, in its framework, there is no longer any conceptual separation between classical and quantum physics. The above equations of motion are both classical, when some of the new terms are absent, and quantum, when all the new terms are present (as we shall see from their integration in terms of a Schr\"odinger equation). However, such a recovered unity of representation does not mean that the quantum and classical realms are equivalent, since the nondifferentiable geometry does have radical effects that are irreducible to classical differentiable physics.

The next step consists of skipping from a classical-type description tool and representation (in terms of geodesic velocity field and equation of dynamics) to the quantum tool and representation (in terms of wavefunction and Sch\"rodinger equation) by merely performing a change of variables. 

The complex wavefunction, $\psi$, which has been introduced in equation~(\ref{cwfunct}) as another expression for the complex action ${\cal S}$, can also been written as
\beq
\psi = \rm{e}^{i{\cal S}/S_0} \; .
\label{cmpwfdef}
\eeq
The real quantity $S_0$, which has been introduced for dimensional reasons, will be given below a physical meaning. From equation~(\ref{vasgrads}), one infers that the function $\psi$ is related to the complex velocity as follows:
\beq
{\cal V} = - i \, \frac{S_0}{m} \, \nabla (\ln \psi) \; .
\label{psiversusv}
\eeq
All the mathematical tools needed to reformulate the fundamental equation of dynamics (equation~(\ref{neweq})) in terms of the new quantity $\psi$ are now available. This equation takes the form
\beq
\nabla   \Phi  =   i S_0 \left[ \frac{\partial }{\partial t} \nabla \ln\psi   - i \left\{  \frac{S_0}{m} (\nabla   \ln\psi  . \nabla   )
(\nabla   \ln\psi ) + {\cal D} \Delta (\nabla   \ln\psi )\right\}\right] \; .
\label{schroderiv1}
\eeq
Using the remarkable identity \cite{LN93,LN07}
\beq
\frac{1}{\alpha} \; \nabla\left(\frac{\Delta  R^{\alpha}}{R^{\alpha}}\right) = 2\alpha \; \nabla \ln R \; . \nabla (\nabla \ln R) + \Delta (\nabla \ln R),
 \label{remid}
\eeq
for $\alpha=S_0/2m{\cal D}$ and $R = \psi$, one obtains, after some calculations,
\beq
\nabla \Phi = i S_0 \, \nabla \left[ \frac{\partial}{\partial t} \ln \psi  - i  \, \frac{2m{\cal D}^2}{S_0} \, \frac{\Delta \psi^{\alpha}} {\psi^{\alpha}} \right] \; .
\label{schroderiv2}
\eeq
Therefore, the right-hand side of the motion equation becomes a gradient whatever the value of $S_0$. Furthermore, equation~(\ref{schroderiv2}) can be written as
\beq
\nabla \Phi = i \, 2m {\cal D} \, \nabla \left[\frac{ \partial{\psi^{\alpha} }/{\partial t} - i  \, {\cal D} \Delta {\psi^{\alpha}}}{{\psi^{\alpha}}} \right] \; ,
\label{schroderiv3}
\eeq
and it can be integrated under the form of a Schr\"odinger equation
\beq
{\cal D}^2 \, \Delta{\psi^{\alpha}} + i {\cal D} \frac{\partial}{\partial t} {\psi^{\alpha}} - \frac{\phi}{2m}{\psi^{\alpha}} = 0 \; ,
\label{genschroeq}
\eeq
Equation~(\ref{cmpwfdef}) defining the wavefunction implies $\psi^{\alpha} = \rm{e}^{i{\cal S}/2m{\cal D}}$. Therefore, to identify $\psi^{\alpha}$ with a wavefunction, $\psi$, such as that defined in equation~(\ref{cmpwfdef}) the constant $S_0$ must verify
\beq
S_0=2m \cal D \; ,
\label{so}
\eeq
which is nothing but a generalized Compton relation, to which a geometric interpretation is now provided. Indeed, the parameter ${\cal D}$ defines the amplitude of the fractal fluctuations through the relation  $\langle d\xi^2 \rangle=2{\cal D}dt$. But one may express this relation in a scale invariant way as $\langle (d\xi/\lambda)^2 \rangle=c dt/\lambda$ by introducing a length scale $\lambda$. Therefore, one finds that the fractal fluctuation parameter defines actually a length scale 
\beq
\lambda= \frac{2{\cal D}}{c} \; ,
\eeq
so that equation~(\ref{so}) becomes $\lambda=S_0/mc$. One recognizes in this relation the Compton relation in which the constant $S_0$ can be fixed by microphysics experiments to the value $\hbar$ and which reads
\beq
\lambda=\frac{\hbar}{mc} \; .
\eeq 
Therefore, equation~(\ref{genschroeq}) finally writes
\beq
\frac{\hbar^2}{2m} \Delta \psi + i \hbar \frac{\partial}{\partial t}\psi = \Phi \psi \; .
\label{schrosr}
\eeq
which is recognized as the time-dependent Schr\"odinger equation for a particle with mass $m$ and wavefunction $\psi$ within a potential $\Phi$. This result reinforces the interpretation of $\psi$ in terms of a wavefunction.

With this general proof, we have therefore derived both a linear Schr\"odinger-type equation of which the complex function ${\psi}$ is a solution and the Compton relation. 

This proof can be further generalized by accounting for all the terms, classical and fractal. This generalization is obtained by simply replacing the classical part of the complex velocity field $\cal V$ by the full velocity field $\widetilde{\cal V}$ at each step of the proof. The successively defined physical quantities (Lagrange function, momentum, action, Hamiltonian, wavefunction) become scale-dependent fractal functions, and in the end one obtains the same Schr\"odinger equation, but which now allows fractal and nondifferentiable solutions. The possible existence of fractal wavefunctions in quantum mechanics had already been suggested by Berry \cite{Berry1996,Hall2004}. It is supported and derived in the scale relativity framework as a very manifestation of the nondifferentiability of spacetime \cite{LN07}.

Note also that a mathematical formulation of the above scale calculus leading to the establishment of the covariant derivative operator (scale operator) and to the derivation of the Schr\"odinger equation in this framework has been performed by Cresson and Ben Adda \cite{BC00,Cresson2001,JC03,JC05}.

\subsection{Hamiltonian approach} \label{hamiltonap}

The more general form of the Schr\"odinger equation, equation~(\ref{sch}), can be recovered in the framework of a Hamilton-like mechanics, as we shall now see. 

First integrate directly the Euler-Lagrange equations as an energy equation expressed in terms of the complex velocity field. For this, start again from the Euler-Lagrange equations of a particle in a scalar potential, equation~(\ref{neweq}), in which the expression of the covariant derivative is expanded to obtain
\beq
m \left( \frac{\partial{\cal V}}{\partial t} + {\cal V}. \nabla {\cal V}-i {\cal D} \Delta {\cal V} \right) = - \nabla  \Phi \; .
\label{newexp}
\eeq
Now, we know that the velocity field ${\cal V}$ is potential, since it writes ${\cal V}= \nabla {\cal S}/m$. Thanks to this property, we have
\beq
m\frac{\partial {\cal V}}{\partial t}= \frac{\partial}{\partial t}( \nabla{\cal S})= \nabla \left( \frac{\partial{\cal S}}{\partial t} \right) \; ,
\eeq
\beq
m{\cal V}.\nabla {\cal V}=\nabla\left( \frac{1}{2} \,m\, {\cal V}^2 \right) \; ,
\eeq
\beq
-i{\cal D}\Delta {\cal V}=-i{\cal D} \nabla(\nabla.{\cal V}) \; .
\eeq
Therefore, all the terms of the motion equation are gradients, so that one obtains, after integration, a prime integral of this equation, up to a constant which can be absorbed in a redefinition of the potential energy $\Phi$. This generalized Hamilton-Jacobi equation writes 
\beq
\frac{\partial{\cal S}}{\partial t}+{\cal H}=0 \; ,
\label{jacobi}
\eeq
with the Hamilton (energy) function given by
\beq
{\cal H}=  \frac{1}{2} \,m\, {\cal V}^2-i m {\cal D} \, \nabla.{\cal V}+ \Phi \; ,
\label{hamilton}
\eeq
i. e., in terms of the complex momentum,
\beq
{\cal H}=  \frac{{\cal P}^2}{2m} -i  {\cal D} \,\nabla.{\cal P}+ \Phi \; .
\eeq
Note that there is an additional potential energy term $-i  {\cal D} \,\nabla.{\cal P}$ in this expression \cite{LN98,JCP99}. This means that the energy is affected by the nondifferentiable and fractal geometry at a very fundamental level, namely, at the level of its initial conceptual definition. This was not unexpected, owing to the fact that the scale relativity approach allows us to propose a foundation of quantum mechanics, while the quantum realm is known since its discovery to have brought radical new features of the energy concept, such as the vacuum energy and its divergence in quantum field theories. 

It is easy to trace back the origin of this term. Namely, while the standard ${\cal V}^2$ term comes from the $ {\cal V}. \nabla$ contribution in the covariant derivative, the additional potential energy $-i m {\cal D} \, {\rm div}{\cal V}$ comes from the second order derivative contribution, $i {\cal D} \Delta$.
A strongly covariant form of the Hamiltonian can be obtained by using the fully covariant form of the covariant derivative operator, (equation~(\ref{OK})). With this tool, the expression of the relation between the complex action and the complex Lagrange function reads
\beq
{\cal L}= \frac{\dfr \,{\cal S}}{dt}= \frac{\d {\cal S}}{\d t} + \widehat{\cal V}. \nabla {\cal S} \; .
\eeq
Since ${\cal P}= \nabla {\cal S}$ and ${\cal H}= -\d {\cal S}/\d t$, one finally obtains for the generalized complex Hamilton function the same form it has in classical mechanics, namely,
\beq
{\cal H}= \widehat{\cal V}. {\cal P} -{\cal L} \; .
\eeq
This is an important new result, which means that a fully covariant form is obtained for the energy equation in terms of the velocity operator $\widehat{\cal V}$.  The additional energy term is recovered from this expression by expanding the velocity operator, which gives ${\cal H}={\cal V}. {\cal P} -i {\cal D} \nabla.{\cal P} -{\cal L}$.

Let us now replace the action in equation~(\ref{jacobi}) by its expression in terms of the wavefunction. We obtain a Hamilton form for the motion equation that reads
\beq
{\cal H} \psi =\left( \frac{1}{2} \,m\, {\cal V}^2-i m {\cal D} \, \nabla.{\cal V}+ \Phi \right) \psi=2i m{\cal D} \,\frac{\partial \psi}{\partial t} \; .
\label{hammotion}
\eeq
However, this is not yet the quantum-mechanical equation, since ${\cal H} \psi$ is a mere product. In order to complete the proof, we need to show that ${\cal H} \psi= \hat{H} \psi$, where $\hat{H}$ is the standard Hamiltonian operator. To this purpose, we replace the complex velocity field by its expression ${\cal V}=-2i {\cal D} \nabla \ln \psi$, and we finally obtain a Schr\"odinger equation under the form
\beq
\hat{H} \psi= 2\,i \,m{\cal D}\, \frac{\partial \psi}{\partial t} \; ,
\eeq
where the Hamiltonian is given by
\beq
\hat{H}= -2m {\cal D}^2 \Delta +\Phi \; ,
\eeq
as expected. By considering the standard quantum-mechanical case $\hbar=2m{\cal D}$, we obtain the usual form of the Schr\"odinger equation,
\beq
i\hbar \, \frac{\partial \psi}{\partial t} = \hat{H} \psi \; .
\label{schroham1}
\eeq

\section{Correspondence principle} \label{corprinc}

When inserting the value $S_0 = \hbar$ in the expressions for ${\cal P}$ and $\psi$ given respectively by equations~(\ref{gencomplmom2}) and (\ref{cmpwfdef}), one obtains
\beq
{\cal P} \psi = -i \hbar \, \nabla \psi \; ,
\label{opmom}
\eeq
which gives, in operator terms,
\beq
\hat{P} = -i \hbar \, \nabla \; .
\label{momop}
\eeq

The Hamiltonian operator has been found in section~\ref{hamiltonap} to be given, in the quantum case where $2 m {\cal D} = \hbar$, by
\beq
\hat{H} = i\hbar \frac{\partial}{\partial t} = - \frac{\hbar^2}{2m}\, \Delta + \Phi \; .
\label{hamop}
\eeq
This implies that the kinetic energy corresponds to the operator
\beq
\hat{T} = -\frac{\hbar^2}{2m}\,\Delta
\label{kinener}
\eeq

As for the position vector $r$, it is present in the Schr\"odinger equation as an operator occurring in the definition of the potential $\Phi(r,t)$ and acting on the wavefunction $\psi$ by multiplying it by its corresponding components $(x,y,z)$ or $(r,\theta, \phi)$, depending on the representation best adapted to the symmetries of the problem.


We have therefore recovered the quantum-mechanical operators under their correct form, including in particular $p^2 \rightarrow -\hbar^2 \Delta$ in the Hamiltonian, which is yet one of the ambiguous cases of standard quantum mechanics \cite{Messiah1959}.

Moreover, as is well known, the operators $-i \hbar \, \nabla$ and $i\hbar \partial / \partial t$ are Hermitian. This is a sufficient condition for these operators to have real eigenvalues. As regards the position operator, it is real, and therefore Hermitian too. These operators are linear and can therefore be shown to operate inside the Hilbert space of the wavefunctions (see, e. g., \cite{CT77}). In standard quantum mechanics, most operator expressions follow from those of $\hat{P}$ and $\hat{R}$, generalizing the correspondence derived for the energy to other dynamical variables.

In standard quantum mechanics, the canonical commutation relations follow straightforwardly from the correspondence principle. The same relations are obtained here. Since the operators `multiply by $r$' and $\nabla$ do not commute, we obtain from equation~(\ref{momop})
\beq
[\hat{R}_i,\hat{P}_j] = i\hbar \delta_{ij} \; .
\label{comrel}
\eeq

\section{The von Neumann and Born postulates} \label{vnbpost}

\subsection{Fluid representation of geodesic equations}\label{froge}

We have given the above two representations of the Euler-Lagrange fundamental equations of dynamics in a fractal and locally irreversible context. The first representation is the  geodesic equation, $\dfr\, {\cal V}/dt=0$, that is written in terms of the complex velocity field, ${\cal V}=V-iU$ and of the covariant derivative operator, $\dfr/dt= \d /dt + {\cal V} . \nabla- i {\cal D} \Delta$. The second representation is the Schr\"odinger equation, whose solution is a wavefunction $\psi$.  Both representations are related by the transformation
\beq
{\cal V}=-2 i{\cal D} \nabla \ln \psi.
\eeq
Let us now write the wavefunction under the form $\psi=\sqrt{P} \times e^{i\theta}$, which amounts to decompose it in terms of a modulus $|\psi|=\sqrt{P}$ and of a phase $\theta$. We shall now build a mixed representation, in terms of the real part of the complex velocity field, $V$, and of the square of the modulus of the wave function, $P=|\psi|^2$.

We separate the real and imaginary parts of the Schr\"odinger equation and make the change of variables from $\psi$, i. e., ($P$, $\theta$), to ($P$, $V$). Thus we obtain respectively a generalized Euler-like equation and a continuity-like equation \cite{LN00--DB52}
\beq
\label{AA1}
\left(\frac{\partial}{\partial t} + V \cdot \nabla\right) V  = -\nabla \left(\frac{\Phi}{m}-2{\cal D}^2 \frac{\Delta \sqrt{P}}{\sqrt{P}}\right),
\eeq
\beq
\label{AA2}
\frac{\partial P}{\partial t} + {\rm div}(P V) = 0.
\eeq
This system of equations is equivalent to the classical system of equations of fluid mechanics (with no vorticity), except for the change from a matter density to a probability density, and for the appearance of an extra potential energy term $Q$ that writes
\beq
\label{Q1}
Q =-2 m {\cal D}^2 \frac{\Delta \sqrt{P}}{\sqrt{P}}. 
\eeq
The existence of this potential energy is, in the scale relativity approach, a very manifestation of the geometry of space, namely, of its nondifferentiability and fractality, in similarity with Newton's potential being a manifestation of curvature in Einstein's general relativity. It is a generalization of the quantum potential of standard quantum mechanics which is recovered in the special case $2m{\cal D}=\hbar$ \cite{EM27,DB52}. However, its nature was misunderstood in this framework, since the variables $V$ and $P$ were constructed from the wavefunction, which is set as one of the axioms of quantum mechanics, such as the Schr\"odinger equation itself. In contrast, in the scale relativity theory, it is from the very beginning of the construction that $V$ represents the velocity field of the fractal geodesics, and the Schr\"odinger equation is derived from the equation of these geodesics.

\subsection{Derivation of the von Neumann postulate}\label{dovnp}

Some measurements are not immediately repeatable, for example when the energy of a particle is measured by noting the length of the track it leaves on a photographic plate while slowing down. In contrast, the measurement of a given component $M_i$ of the magnetic moment of an atom in a Stern-Gerlach experiment can be repeated immediately by passing the beam through another apparatus. In this case, we expect that if a particular value $m_n$ has been obtained for the first measurement, the same value will be obtained in the second one. This is actually what is observed. Since the result of the second measurement can be predicted with certainty, von Neumann \cite{JVN55} has stated as a postulate of quantum mechanics that, after a measurement, the state of the system is described by the eigenfunction $\psi_n$ of $M_i$ corresponding to the eigenvalue $m_n$.

In section~\ref{cstatef} we have identified the wave particle with the various geometric properties (in position and scale space) of a subset of the fractal geodesics of a nondifferentiable space. This includes the properties of quantization, which are at the origin of the quantum and particle view of quantum mechanics and which are a consequence of the properties of its equations. In such an interpretation, a measurement (and more generally any knowledge acquired about the system, even not linked to an actual measurement) amounts to a selection of the sub-sample of the geodesic family in which only the geodesics having the geometric properties corresponding to the measurement result are kept. Therefore, just after the measurement, the system is in the state given by this result, in accordance with von Neumann's postulate of quantum mechanics. 

Such a state is described by the wavefunction or state function $\psi$, which, in the present approach, is a manifestation of the various geometric properties of the geodesic fluid, whose velocity field is ${\cal V}= -2i { \cal D} \nabla \ln \psi$. Now, one should be cautious about the meaning of this selection process. It means that a measurement or a knowledge of a given state is understood, in the scale relativity framework, as corresponding to a set of geodesics which are characterized by common and definite geometric properties. Among all the possible virtual sets of geodesics of a nondifferentiable space(time), these are therefore `selected'. But it does not mean that these geodesics, with their geometric properties, were existing before the measurement process. Indeed, it is quite possible that the interaction involved in this process be at the origin of the geometric characteristics of the geodesics, as identified by the measurement result. Scale relativity does not involve a given, static space with given geodesics, but instead a dynamic and changing space whose geodesics are themselves dynamic and changing. In particular, any interaction and therefore any measurement participates in the definition of the space and of its geodesics.

The situation here is even more radical than in general relativity. Indeed, in Einstein's theory, the concept of test particle can still be used. For example, one may consider a static space such as given by the Schwarzschild metric around an active gravitational mass $M$. Then the equation of motion of a test particle of inertial mass $m\ll M$ depends only on the active mass $M$ which enters the Christoffel symbols and therefore the covariant derivative. This is expressed by saying that the active mass $M$ has curved spacetime and that the test particle follows the geodesics of this curved spacetime. Now, when $m$ can no longer be considered as small with respect to $M$, one falls into a two-body problem which becomes very intricated. Indeed, the motion of the bodies enters the stress-energy tensor, so that the problem is looped. The general solutions of Einstein's equations become extremely complicated in this case and are therefore unknown in an exact way. 

However, in scale relativity, even the one-body problem is looped. It is the inertial mass of the `particle' itself whose motion equation is searched for,that enters the covariant derivative. This is indeed expected of a microscopic description of a space(time) which is at the level of its own objects, and in which, finally, one cannot separate what is `space' (the container) from what is the `object' (contained).  In this case the geometry of space and therefore of the geodesics is expected to continuously evolve during the time evolution and also to depend on the resolution at which they are considered.

The von Neumann postulate is a direct consequence of the identification of a `particle' or of an ensemble of `particles' with families of geodesics of a fractal and nondifferentiable spacetime. Namely, a quantized (or not quantized) energy or momentum is considered as a non-local conservative geometric property of the geodesic fluid. Now, in a detection process the particle view seems to be supported by the localisation and unicity of the detection. But we consider that this may be a very consequence of the quantization (which prevents a splitting of the energy) and of the interaction process needed for the detection. For example, in a photon by photon experiment, the detection of a photon on a screen means that it has been absorbed by an electron of an atom of the screen, and that the geodesics are therefore concentrated in a zone of the order of the size of the atom.

Any measurement, interaction or simply knowledge about the system can be attributed to the geodesics themselves. In other words, the more general geodesic set which served to the description of the system before the measurement or knowledge acquisition (possibly without interacting with the system from the view point of the variables considered) is instantaneously reduced to the geodesic sub-set which corresponds to the new state. For example, the various results of a two-slit experiment -- one or two slits opened, detection behind a slit, detection by spin-flip that does not interact with the position and momentum which yields a pure `which-way information', quantum eraser, etc \cite{Feynman1964,Scully1991} -- can be recovered in the geodesic interpretation \cite{LN93,LN96}.

\subsection{Derivation of the Born postulate}\label{dotbp}

As a consequence of the fluid interpretation, the probability for the `particle' to be found at a given position must be proportional to the density of the geodesic fluid. We already know its velocity field, whose real part is given by $V$, identified, at the classical limit, with a classical velocity field. But the geodesic density $\varrho$ has not yet been introduced. In fact, we are, as regards the order of the derivation of the various `postulates', in the same situation as during the historical construction of quantum mechanics. Indeed, the Born interpretation of the wavefunction as being complex instead of real and such that $P=|\psi|^2$ gives the presence probability of the particle \cite{Born1926,Born1927} has been definitively settled after the establishment of the other postulates. Here, we have been able to derive the existence of a wavefunction, the correspondence principle for momentum and energy and the Schr\"odinger equation without using a probability density. 

Now, we expect the geodesic fluid to be more concentrated at some places and less at others, to fill some regions and to be nearly vanishing in others, as does a real fluid. This behavior should be described by a probability density of the paths. However, these paths are not trajectories. They do exist as geometrical `objects', but not as material objects. As an example, the geodesical line between two points on the terrestrial sphere is a great circle. It does exist as a virtual path, whether it is followed by a mobile or not.

The idea is the same here. The geodesic fluid is defined in a purely geometrical way as an ensemble of virtual paths, not of real trajectories. Now, in a real experiment, one may emit zero, one, two, or a very large number of `particles', under the conditions virtually described by the geometric characteristics of space and of its geodesics, namely, fractal-nonfractal transition that yields the mass, initial and possibly final conditions (in a probability amplitude-like description \`a la Feynman), limiting conditions, etc.

When the `particle' number is small, the fluid density manifests itself in terms of a probability density, as in particle by particle two-slit experiments. When the `particle' number is very large, it must manifest itself as a continuous intensity (as the intensity of light in a two-slit experiment). But since, originally, the interferences are those of the complex fluid of geodesics (in a two-slit experiment when both slits are open), we expect them to exist even in the zero particle case. This could be interpreted as the origin of vacuum energy.

In order to calculate the probability density, we remark that it is expected to be a solution of a fluid-like Euler and continuity system of equations, namely,
\beq
\label{AAA1}
\left(\frac{\partial}{\partial t} + V \cdot \nabla\right) V = - \nabla\left({\frac{\Phi}{m}+Q}\right) \; ,
\eeq
\beq
\label{AAA2}
\frac{\partial \varrho}{\partial t} + {\rm div}(\varrho V) = 0 \; ,
\eeq
where $\Phi$ describes an external scalar potential possibly acting on the fluid, and $Q$ is the potential that is expected to possibly appear as a manifestation of the fractal geometry of space (in analogy with general relativity). This is a system of four equations for four unknowns, $(\varrho,V_x,V_y,V_z)$. The properties of the fluid are thus completely determined by such a system.

Now these equations are exactly the same as equations~(\ref{AA1}) and (\ref{AA2}), except for the replacement of the square of the modulus of the wavefunction $P=|\psi|^2$ by the fluid density $\varrho$. Therefore this result allows us to univoquely identify $P$ with the geodesic probability density, i. e., with the presence probability of the `particle' \cite{CN04,LN07}. Moreover, we identify the non-classical term $Q$ with the new potential which emerges from the fractal geometry. Numerical simulations, in which the expected probability density can be obtained directly from the geodesic distribution without writing the Schr\"odinger equation, support this result \cite{RH97}.

We have seen in section~\ref{post} that this interpretation requires that the sum of the contributions $|\psi|^2$ for all values of $(r,s)$ at time $t$ be finite, i. e., the physically acceptable wavefunctions are square integrable. Now, the linearity of the Schr\"odinger equation implies that if two wave functions $\psi_1$ and $\psi_2$ are solutions of this equation for a given system, $c_1\psi_1 + c_2\psi_2$, where $c_1$ and $c_2$ are any complex numbers, is also a solution. From both these properties we can define, such as done in the standard framework of quantum physics, the space of the wavefunctions of a given physical system which can be provided with the structure of a vectorial Hilbert space (see, e. g., \cite{CT77}).

Remark that the above proof of the Born postulate relies on the general conditions under which one can transform the Schr\"odinger equation into a continuity and fluid-like Euler system. By examining these conditions in more detail it is easy to verify that it depends heavily on the form of the free particle energy term in the Lagrange function, $p^2/2m$. In quantum-mechanical terms, this means that it relies on the free kinetic energy part of the Hamiltonian, $\widehat{T}=-(\hbar^2/2m)\Delta$. In particular, one finds that the quantum potential writes in terms of this operator
\beq
Q= \frac{\widehat{T}\sqrt{P}}{\sqrt{P}}.
\eeq
As we have argued in section~\ref{corprinc}, this form of the work operator is very general since it comes from motion relativity itself (in the non-relativistic limit $v \ll c$ that is considered here). However, if there existed some (possibly effective) situations in which this form would no longer be true, the derivation of equations~(\ref{AA1}) and (\ref{AA2}) would no longer be ensured and the Born postulate would no longer hold in the scale relativistic framework. This could yield a difference between theoretical predictions of the scale relativity theory and those of standard quantum mechanics, for which the Born postulate $P=|\psi|^2$ is always true, and therefore this could provide us with a possibility to put both theories to the test.

\section{Miscellaneous issues} \label{miscel}

\subsection{Differentiable or nondifferentiable wavefunction} \label{dndwf}

We have seen in section~\ref{post} that one of the properties initially postulated for the wavefunction is that it should be differentiable. The linear second order differential Schr\"odinger equation established in section~\ref{schroeq} admits of course differentiable solutions. Therefore the postulated property is recovered for the wavefunction $\psi$. 

However, we have seen that scale relativity allows one to derive a generalized Schr\"odinger equation whose solutions can be nondifferentiable (they are described as explicitly scale-dependent fractal functions that diverge when the scale interval tends to zero), therefore extending the possible application domain of quantum mechanics and possibly providing future interesting laboratory tests of the predictions of the new theory. Such a result agrees with Berry's \cite{Berry1996} and Hall's \cite{Hall2004} similar findings obtained in the framework of standard quantum mechanics. 

\subsection{Spin, charges and relativistic quantum mechanics} \label{spin}

In the above derivations, we have, for simplification purpose, limited ourselves to the non-relativistic case (in the sense of motion relativity) and to the spin-less case.

However, the tools that lead to derive the various postulates of non-relativistic quantum mechanics from the geodesics of a fractal space can easily be generalized to the relativistic case, which amounts to consider a full fractal spacetime \cite{LN96,CN04,LN94}. One derives in this case the Klein-Gordon equation and, more generally, the Dirac equation \cite{Ord1983,CN04}.
 
We have indeed demonstrated in previous works \cite{CN04,CN06} that spin arises naturally in the framework of scale relativity when one takes into account, in addition to the symmetry breaking under the reflexion $dt \leftrightarrow -dt$, a new symmetry breaking under the discrete transformation $dx^\mu \leftrightarrow - dx^\mu$, which also emerges as a direct consequence of nondifferentiability. Moreover, the standard quantum-mechanical properties of spin are thus recovered. In particular, its eigenvalues for the electron, $\pm \hbar/2$, proceed from the bi-spinorial nature of its wavefunction, which is a consequence of the additional two valuedness of the velocity field of geodesics that comes from the new discrete symmetry breaking. Then one finds that the geodesic equation for this velocity field can be integrated under the form of the Dirac equation \cite{CN04}. The Pauli equation has also been derived in the scale relativity framework and, even though it comes under the non-relativistic part of the theory, we have shown that, as in standard quantum mechanics, the origin of spin in this equation is fundamentally relativistic \cite{CN06}.

Electromagnetic (Abelian) \cite{LN94} and non-Abelian fields and charges may also be founded in a geometric way on the principles of scale relativity. We refer the interested reader to \cite{NCL06}.

\subsection{Quantum mechanics of many identical particles} \label{many}

The fractal and nondifferentiable space(time) approach of scale relativity allows one to recover easily the many identical particle Schr\"odinger equation and to understand the meaning of the non-separability of particles in the wavefunction \cite{LN96}. The key to this problem lies clearly in the fact that the wavefunction is nothing but another expression for the action, itself being complex as a consequence of the nondifferentiable geometry of space.

The fact that the motion of $n$ particles in quantum mechanics is irreducible to their classical motion, i. e., it cannot be understood as the sum of $n$ individual motions but must be instead taken as a whole, agrees with our identification of `particles' with the geometric properties of virtual geodesics in a fractal space(time). Being reduced to these geometrical properties, identical `particles' become totally indiscernible. Since, in this framework, they are nothing but an ensemble of purely geometrical fractal `lines', there is absolutely no existing property that could allow one to distinguish between them.

Recall, moreover, that these lines should not be considered in the same way as standard nonfractal curves. They are defined in a scale relative way, i. e., only the ratio from a finite scale to another finite scale does have meaning, while the zero limit, $dt \to 0$ and $dx \to 0$, is considered to be undefined. As a consequence, the concept of a unique geodesical line loses its physical meaning. Whatever be the manner to unfold them or to select them and whatever the scale, the definition of a `particle' always involves an infinity of geodesics. Now, it is also clear that the ensemble of geodesics that describes, e. g., two `particles', is  globally different from a simple sum of the geodesic ensemble of the individual `particles'. This is properly described by the wavefunction $\psi(x_1,x_2)$ (where $x_1$ and $x_2$ are the position vectors of the two `particles'), which is the solution of the Schr\"odinger equation, from which one can recover the velocity fields of the `particles', ${\cal V}_1=-i (\hbar/m) \nabla_1 \ln \psi$ and ${\cal V}_2=-i (\hbar/m) \nabla_2 \ln \psi$.

\section{Discussion} \label{disc}

In the first part of this paper (section~\ref{post}), we have proposed a classification of the different statements which can be found in the literature as postulates of quantum mechanics. We retain five `main' postulates as being the fundamental assumptions from which `secondary' postulates can be demonstrated. Three derived principles are also recalled for completeness.

Then we have shown that the `main' postulates can be derived from the founding principles of scale relativity. The complex nature of the wavefunction is issued from the twovaluedness of the velocity field which itself is a consequence of space(time) nondifferentiability. The Schr\"odinger equation follows as being the integral of a geodesic equation in a fractal space. The correspondence principle is derived from the covariance of the construction and the basic symmetries which therefore hold in this framework. Then, we have recovered Born's postulate from the fluid-like nature of the velocity field of geodesics. Von Neumann's postulate and its corollary reduction of the wavefunction is also a consequence of the nature of the geodesic fluid and of the identification of the `particle' with the various geometric properties of the fractal geodesics.

To complete this reconstruction of the quantum-mechanical bases, we have chosen to work within the coordinate realization of the state function, since it is in this representation that the scale relativistic formulation is the most straightforward. However, since we have demonstrated that the wavefunctions can be defined as vectors of an Hilbert space and that to each classical dynamical quantity can be attached an Hermitian operator acting in such a space, the standard quantum-mechanical formalism is recovered and a switch to momentum space as well as to a Dirac representation can be implemented by the same methods as those used in this standard framework.

For readability purpose, we have also limited, in this paper, our consideration to operators which are constructed with the position and momentum ones and we have referred the reader in section~\ref{spin} to previous works for the study of other internal or external observables such as those yielding spin, electromagnetic and non-Abelian charges.

We want finally to stress that, owing to the fluid-like interpretation pertaining to the scale relativity space(time) description, any geodesic bundle of such a fractal space(time) relating any couple of its points (or starting from any point) always contains an infinity of geodesics and fills space like a fluid, therefore allowing one to directly relate the velocity field of this geodesic fluid to the wavefunction. Moreover, each one of the  `individual' geodesical curves participating in this bundle is itself resolution dependent and is actually undefined at the limit $dt \to 0$ and $dx \to 0$, i. e., it is a fractal curve irreducible to a standard curve of vanishing thickness (section~\ref{vnbpost}). These two non-classical features show that the scale relativistic description owns at a fundamental level the properties of non-locality and unseparability which underlie quantum-mechanical specific features such as entanglement. Moreover, since the ultimate fractal space(time), for which the resolution $dt$ vanishes, is unreachable (and therefore the complete distribution of the stochastic variables $d \xi$ is unknown, only the two first moments are defined), this precludes the possibility of considering the scale relativistic construction of quantum mechanics as a hidden parameter theory.

Actually, such a description in terms of an infinity of fractal and nondifferentiable paths is a continuation of Feynman's attempts to come back to a spacetime approach to quantum mechanics, which culminated in his path integral formulation \cite{RF48}. Such a filiation is also claimed by Ord in his own fractal spacetime description \cite{Ord1983}. Feynman's path integral decomposes the wavefunction in terms of the sum over all possible paths of equiprobable elementary wavefunction $e^{iS_{\rm cl}}$, where $S_{\rm cl}$ is the classical action. As explained by Feynman and Hibbs, `the sum over paths is defined as a limit, in which at first the path is specified by giving only its coordinate $x$ at a large number of specified times separated by very small intervals $\varepsilon$. The path sum is then an integral over all these specific coordinates. Then to achieve the correct measure, the limit is taken as $\varepsilon$ approaches zero' (\cite{FH65}, p. 33). It therefore involves explicitly in its estimate the use of a scale variable $\varepsilon$, which is just the basic method of scale relativity. Now, most of these paths are nondifferentiable, and the same is true of those which contribute mainly to the path integral (as recalled in the introduction). Indeed, the problem with the standard path integral approach is that some of the fundamental properties of the paths diverge when taking the limit $\varepsilon \to 0$, in particular the mean-square velocity which reads $-\hbar/(i m \Delta t)$ (\cite{FH65}, p. 177). In scale relativity the problem is solved by simply not taking the limit (limit which is actually physically meaningless) and by defining the mean-square velocity as an explicitly scale-dependent fractal function of $\Delta t$. This allows us to work with quantities that are infinite in the standard sense, then to go one step deeper, and to consider these paths, no longer as fractal and nondifferentiable trajectories in standard spacetime, but as geodesics of a spacetime which itself is nondifferentiable and fractal. Particles can then be defined by the geometry of the paths themselves. One no longer needs to carry anything along them (not even an elementary wavefunction with unit probability), since the whole wavefunction, including its modulus and its phase, can be constructed from the flow of geodesics itself.

Let us now continue the discussion by comparing the scale relativity approach with other approaches to the foundation of quantum mechanics. In this respect, we remark that the various attempts that have been made to interpret it while keeping classical concepts are unsatisfactory. Four of these attempts are worth being examined in the present context.

In the `quantum potential' approach of de Broglie \cite{LB26} and Bohm \cite{DB52} the particle trajectories are deterministic and the statistical behaviour is artificially introduced by assuming that the initial conditions are at random. The physical origin of the `quantum force' that produces quantum effects remains unexplained.

In `stochastic quantum mechanics' \cite{RF33--EN66}, it is assumed that an underlying Brownian motion, of unknown origin, is at work on every particle. This Brownian force induces a Wiener-like process which is at the origin of the quantum behaviour. To derive the Schr\"odinger equation, Nelson \cite{EN66} defines two diffusion processes yielding a so-called `forward Fokker-Planck equation' and a `backward  Fokker-Planck equation'. However, it has been shown that the `backward  Fokker-Planck equation', which is set as a founding equation for stochastic quantum mechanics, does not actually correspond to any known classical process \cite{GH79}. Moreover, it has also been shown that the quantum-mechanical multitime correlations do not agree, in general, with the stochastic-mechanical multitime correlations when the wavefunction reduction is taken into account \cite{WL93}. Both demonstrations preclude the possibility of obtaining a classical diffusion theory of quantum mechanics.

In `geometric quantum mechanics' \cite{ES84,CC91}, the quantum behaviour is found to be a consequence of an underlying Weyl geometry, and the statistical description is obtained by the same postulate of random initial conditions as that pertaining to the `quantum potential' approach. Then `the theory does not describe the motion of an individual particle; rather it describes the statistical behaviour of an ensemble of identical particles' \cite{ES84}, so that it is not able to yield a probabilistic description emerging from the processes affecting individual systems, as demanded by Einstein \cite{AP82}. As shown by Castro \cite{CC90}, the Bohm quantum potential equation is recovered, rather than set, from a least-action principle acting on the Weyl gauge potential. Such an approach may be {\it partly} related to ours, since Weyl's geometry is closely related to conformal transformations which include dilations that are part of the scale transformation set. But this geometry and its non-Abelian generalization, once it is reinterpreted in the scale relativity framework, is expected to act at the level of the emergence of gauge fields rather than of the quantum behavior itself \cite{NCL06}.

More recently, Adler and coworkers (see \cite{SA04} and references therein, and \cite{SA06}) developed the idea that quantum mechanics might be an emergent phenomenon arising from the statistical mechanics of matrix models which have a global unitary invariance. This proposal, called `trace mechanics', leads to the usual canonical commutation/anticommutation algebra of quantum mechanics, as well as to the Heisenberg time evolution of operators, which in turn imply the usual rules of the Schr\"odinger picture. However, as stressed by the proponents themselves, they are not able at this point to give {\it the} specific theory that obeys all the required conditions. They even contemplate the possibility that there might be no trace-mechanical theory satisfying all the quantum-mechanical prescriptions. We must therefore consider this `theory' as tentative and wait for possible improved developments.

Another related subject worth to be discussed in this context is the question of the mathematical tools that could be best adapted to study possible generalizations of quantum-mechanical laws. The scale relativity tools allowing to found the quantum mechanics postulates are fundamentally geometrical and relativistic in their essence: namely, the basic concepts are nondifferentiable spacetime, geodesics and covariant derivatives. Now, in order to implement these purely geometric concepts, we have been naturally led to introduce complex numbers \cite{LN93}, then quaternionic and biquaternionic numbers \cite{CN04,CN06}, and then multiplets of biquaternions \cite{NCL06}. This introduction comes as a mere simplifying representation of the successive two valuednesses of variables which are a consequence of the nondifferentiable geometry itself. This relates the scale relativity approach to other various attempts of generalizations of the mathematical tool of quantum mechanics, in particular by the use of Clifford algebras.

Hence, following a proposal first put forward by Nottale \cite{LN97A}, Castro and Mahecha \cite{CM06} adopted a complex-valued constant ${\cal D}$, which they interpreted in the quantum realm as a complex reduced Planck constant $\hbar$. Using the formalism of scale relativity, they obtain a new nonlinear Schr\"odinger equation, where the potential and the reduced Planck constant are complex. Despite having non-Hermitian Hamiltonian, contrary to what we have obtained above with a real-valued $\hbar$, they still can have eigenfunctions with real-valued energies and momenta, like plane waves and soliton solutions. Such an approach is faced with the problem of finding a geometric interpretation of a complex diffusion coefficient and of a complex Planck constant, and it therefore awaits a physical domain of application; otherwise it may be considered as an ad-hoc mathematical generalization \cite{JC03}. 

In this context, it should be stressed again that, contrary to stochastic approaches, in scale relativity, the fractal geodesics are not trajectories followed by `particles' (which would own internal properties such as mass, spin, charge, etc), but they are purely geometric paths (fractal and in infinite number) from which the wave particle's conservative quantities emerge. For example, the mass is a manifestation of the fractal fluctuations, namely, $m=\hbar dt/<\!\! d \xi^2 \!\!>$; the energy-momentum is given by the transition from scale dependence to effective scale independence in scale space, identified with the de Broglie scale, namely, $p^\mu= \hbar/\lambda^\mu$; the spin is an intrinsic angular momentum of the fractal geodesics \cite{CN06}; the charges are the conservative quantities that appear, according to Noether's theorem, from their internal scale symmetries \cite{LN96,NCL06}. The existence of quantas, upon which the concept of `particle' is based, is just a consequence of the quantization of these geometric properties, which itself is derived from the properties of the geodesic equations (which can be integrated in terms of the standard quantum-mechanical equations after change of variable). Furthermore, the scale-dependent part of the elementary displacement is described by a stochastic variable which is not mandatorily Gaussian. This is just the strength of the scale relativity approach that the description of the fractal fluctuations is fully general and needs no special hypothesis for recovering the quantum-mechanical tools and equations. Thus quantum mechanics appears, in this framework, as a general manifestation of any kind of nondifferentiable spacetime, not of a particular one.

In another work, still following a proposal by Nottale \cite{LN97A} consisting of replacing the real-valued constant ${\cal D}$ by a tensor ${\cal D}_{jk}$ (which still allows us to obtain a Schr\"odinger equation of a generalized form) Castro \cite{CC06} derived new Klein-Gordon and Dirac equations with a matrix-valued extension of Planck's constant in Clifford space. However, the results of this work yield modified products of (Heisenberg) uncertainties and modified dispersion relations in contradistinction to what occurs in ordinary quantum mechanics. According to the author himself, these results might only apply at very small scales, near the Planck scale, e. g., in the framework of a quantum theory of gravity. This is very far from our own purpose which is here to recover the axioms of standard quantum mechanics from physical first principles, in a way which is expected to be valid at the accessible scales of molecular, atomic, nuclear and elementary particle physics.

Another work by Adler \cite{SA95} gives a systematic development of quaternionic quantum mechanics paralleling the standard textbook treatments of complex mechanics, with the wavefunction complex values replaced by quaternionic ones. His purpose is twofold: to get a better understanding of the theory by adding a new mathematical realization to the ones already available and to aim at a successful unification of the fundamental forces which might require a generalization beyond complex mechanics. However, differently from our scheme, Adler did not derive the axioms of quantum mechanics from his reasoning, but actually set these axioms as his founding principles and showed that the quaternions were compatible with them to justify their use.

\section{Conclusion} \label{concl}

Finally, we remark that, as regards the scale relativity theory and despite its successes, we cannot yet claim that it allows us to found the true quantum mechanics which is really experimented in nature. This would be actually impossible (and this impossibility also applies to the present axiomatic theory of quantum mechanics), since all we can do is to attempt to falsify the theory by experiments. Moreover, the fact that we recover the postulates of standard quantum mechanics does not prove that some other consequences of the founding principles of the scale relativity theory could not contradict them. If such inconsistencies were to appear, either they could be at variance with already known experimental results, in which case they would rule out the theory in its present form, or, more interestingly, they could be used to put the new theory (and also standard quantum mechanics) to the test. 

Let us conclude by listing some original features of the scale relativity approach that have been pointed out in this paper or in other complementary works (see the review papers \cite{LN04,LN07B}), and which could lead to such experimental tests in the future: (i) nondifferentiable and fractal solutions of the Schr\"odinger equation; (ii) zero particle interference in a Young slit experiment; (iii) breaking of the Born postulate for a possible effective kinetic energy operator $\widehat{T}\neq-(\hbar^2/2m)\Delta$; (iv) underlying quantum regime in the classical domain, at scales far larger than the de Broglie scale; (v) macroscopic systems described by a quantum-type mechanics based on a generalized parameter ${\cal D} \neq \hbar/2m$; (vi) quantum-mechanical systems with a fractal dimension different from $D_F = 2$.



\end{document}